\documentclass[12pt]{article}

\usepackage[T2A]{fontenc}
\usepackage[cp1251]{inputenc}
\usepackage[russian]{babel}

\usepackage{amsmath}
\usepackage{a4wide}
\usepackage{amsfonts,amssymb}
\usepackage{graphicx}

\begin{document}

\noindent {\sl Problems of Information Transmission},\\
\noindent vol. 51, no. 4, pp. 3-22, 2015.

\vskip 0.8cm
 
\begin{center} {\bf M. V. Burnashev}
\end{center}

\begin{center}
{\large\bf ON RELIABILITY FUNCTION OF BSC: EXPANDING THE REGION,
WHERE IT IS KNOWN EXACTLY
\footnote[1]{Supported in part by the Russian Foundation for
Basic Research (project nos. 15-01-08051) and by Russian Scientific
Foundation (project nos. 13-01-12458 офи\_м2).}}
\end{center}

The region of rates (``straight-line''), where the BSC reliability
function is known exactly, is expanded.

\medskip

\begin{center}
{\large\bf \S\;1. Introduction and main results}
\end{center}

In the paper notation from \cite{Bur6} is used. A binary symmetric
channel (BSC) with crossover probability $0 <p < 1/2$ and $q=1-p$
is considered. Let $F^{n}$ denote the set of all $2^{n}$ binary
$n$-tuples, and $d(\boldsymbol{x},\boldsymbol{y})$,
$\boldsymbol{x},\boldsymbol{y} \in F^{n}$
denote the Hamming distance between  $\boldsymbol{x}$ and
$\boldsymbol{y}$. A subset ${\cal C} = {\cal C}(M,n) =
\{\boldsymbol{x}_{1},\ldots,\boldsymbol{x}_{M}\} \subseteq F^{n}$ is
called a {\it code of length} $n$ and {\it cardinality} $M$.
The {\it minimum distance} of the code ${\cal C}$ is
$d({\cal C}) = \min \{d(\boldsymbol{x}_{i},\boldsymbol x_{j}):
i \neq j\}$. Code rate is $R({\cal C}) = n^{-1}\log_{2}M$.
Everywhere below $\log z = \log_{2}z$.

Cardinality of a set $A$ is denoted by $|A|$. Code {\it spectrum}
({\it distance distribution})
$B({\cal C}) = (B_{0},B_{1},\ldots,B_{n})$
is the $(n+1)$--tuple with components
\begin{equation}\label{defBi}
B_{i} = \left|{\cal C}\right|^{-1}\left|\left\{
(\boldsymbol{x},\boldsymbol{y}): \boldsymbol{x},\boldsymbol{y}
\in {\cal C},\,d(\boldsymbol x,\boldsymbol{y}) = i\right\}\right|,
\qquad i=0,1,\ldots,n.
\end{equation}
In other words, $B_{i}$ is average number of codewords
$\boldsymbol{y}$ on the distance $i$ from the codeword
$\boldsymbol x$. Clearly, $B_{0} + \ldots + B_{n} = |{\cal C}|$.
The total number of ordered codepairs
$\boldsymbol{x},\boldsymbol{y} \in {\cal C}$ with
$d(\boldsymbol x,\boldsymbol{y}) = i$ equals $|{\cal C}|B_{i}$.

The BSC {\it reliability function} $E(R,p)$ is defined as follows
\cite{E, SGB1, G1}
$$
E(R,p) = \limsup_{n \to \infty}\,\frac{1}{n},
\ln \frac{1}{P_{e}(R,n,p)},
$$
where $P_{e}(R,n,p)$ -- the minimal possible decoding error
probability $P_e$ for $(n,R)$--code.

Introduce the function \cite{M4}
\begin{equation}\label{defG}
G(\alpha,\tau) = 2\frac{\alpha(1-\alpha) - \tau(1-\tau)}
{1+2\sqrt{\tau(1-\tau)}} = \frac{1}{2} - \sqrt{\tau(1-\tau)} -
\frac{(1-2\alpha)^{2}}{2\left[1+2\sqrt{\tau(1-\tau)}\right]}
\end{equation}
and define the function $\delta_{GV}(R) \leq 1/2$
(Gilbert- Varshamov bound) as
\begin{equation}\label{defGV}
1 - R = h_{2}(\delta_{GV}(R)), \qquad 0 \leq R \leq 1,
\end{equation}
where $h_{2}(x) =-x\log_{2}x - (1-x)\log_{2}(1-x)$.

Define the value $R = R(\alpha,\tau)$ by the formula
\begin{equation}\label{defR}
R = 1 - h_{2}(\alpha) + h_{2}(\tau), \qquad
0 \leq \tau \leq \alpha \leq 1/2.
\end{equation}
For $R \in (0,1)$ and $\alpha \in [\delta_{GV}(R),1/2]$ introduce
also values
\begin{equation}\label{defGR11}
\begin{gathered}
\tau_{R}(\alpha) = h_{2}^{-1}\left(h_{2}(\alpha)-1+R\right) \leq
\alpha, \\
\omega_{R}(\alpha) = G(\alpha,\tau_{R}(\alpha))
\end{gathered}
\end{equation}
and
\begin{equation}\label{defGR}
\begin{gathered}
\omega_{R} = \min_{\delta_{GV}(R) \leq \alpha \leq 1/2}
\omega_{R}(\alpha) = \min \{G(\alpha,\tau):
h_{2}(\alpha) - h_{2}(\tau) = 1 - R\}.
\end{gathered}
\end{equation}
The best known upperbound for the maximal relative code distance
$\delta(R)$ (linear \\ programming bound) has the form \cite{M4, Lev1}
\begin{equation}\label{MRRW}
\delta(R) \leq \delta_{LP}(R) = \omega_{R}.
\end{equation}

If $R \leq R_{0} \approx 0.30524$ ($R_{0}$ is defined in
\eqref{defR0} and \eqref{eqR0}), then minimum in the right-hand side
of \eqref{defGR} is attained for $\alpha =1/2$, and the formula
\eqref{MRRW} takes simple form \cite{M4, Lev1}
\begin{equation}\label{F01}
\begin{gathered}
\delta_{LP}(R) = \omega_{R} = \frac{1}{2} - \sqrt{\tau(1-\tau)},
\qquad R = h_{2}(\tau).
\end{gathered}
\end{equation}

Denote by $\alpha_{R},\tau_{R}$ optimal values of parameters
$\alpha,\tau$ в \eqref{defGR},i.e.
\begin{equation}\label{defalphaR}
\begin{gathered}
\omega_{R} = \omega_{R}(\alpha_{R}) = G(\alpha_{R},\tau_{R}).
\end{gathered}
\end{equation}
The function $\omega_{R}$ decreases monotonically for $R \in (0,1)$,
and the function $\alpha_{R}$ does not increases in $R$.

Introduce critical rates $R_{\rm crit}(p), R_{1}(p)$ and
$R_{2}(p)$, beginning with well-known
\begin{equation}\label{defRcrit}
R_{\rm crit}(p) =
1 - h_{2}\left(\frac{\sqrt{p}}{\sqrt{p}+\sqrt{q}}\right).
\end{equation}

The rate $R_{1}(p)$ was introduced in \cite[формулы (6)]{Bur6}
\begin{equation}\label{defR1}
\begin{gathered}
R_{1}(p) = h_{2}(\tau_{1}(p)), \qquad \tau_{1}(p) =
\frac{\left(1-(4pq)^{1/4}\right)^{2}}{2(1+\sqrt{4pq})} \leq
\frac{1}{2}.
\end{gathered}
\end{equation}

Introduce the important value
\begin{equation}\label{defomega1}
\begin{gathered}
\omega_{1}(p) = \frac{2\sqrt{pq}}{1 + 2\sqrt{pq}} =
G(1/2,\tau_{1}(p)), \qquad 0 \leq p \leq 1/2.
\end{gathered}
\end{equation}
Note that the value $\omega_{1}(p)$ is defined by the condition
$t_{1}(\omega) = t_{2}(\omega,p)$ (см. \eqref{S.3}).

Introduce the critical rate $R_{2} = R_{2}(p)$ by the formula
\begin{equation}\label{defR2}
\begin{gathered}
\omega_{R_{2}} = \omega_{1}(p)= \frac{2\sqrt{pq}}{1 + 2\sqrt{pq}},
\qquad 0 < p < 1/2,
\end{gathered}
\end{equation}
or, equivalently,
\begin{equation}\label{syst2a}
R_{2}(p) = 1 - \max_{\alpha,\tau}\{h_{2}(\alpha)- h_{2}(\tau):
G(\alpha,\tau) = \omega_{1}(p)\}.
\end{equation}
In other words, $R_{2}(p)$ is the minimal rate for which it is
possible to have $G(\alpha,\tau) = \omega_{1}(p)$. On the contrary,
$R_{1}(p)$ is the maximal such rate (it corresponds to
$\alpha = 1/2$). Functions $R = R_{2}(p)$ and $R = R_{1}(p)$
decreases monotonically in $p \in [0,1/2]$.

Introduce the value  $p_{0} \approx 0,036587$ as the unique root of
the equation $R_{2}(p) = R_{1}(p)$. If $p < p_{0}$, then in
the optimizing value $\alpha < 1/2$. If $p \geq p_{0}$, then
the optimizing value $\alpha = 1/2$ and $R_{2}(p) = R_{1}(p)$.
We also have $R_{2}(0) = R_{1}(0) = 1$ and
$R_{2}(p_{0}) = R_{1}(p_{0}) = R_{0} \approx 0.30524$, where
$R_{0}$ is defined in \eqref{defR0}--\eqref{eqR0}.

Introduce also the value $p_{1} \approx 0,0078176$ as the unique root
of the equation $R_{1}(p) = R_{\rm crit}(p)$. Then we have
\begin{equation}\label{F011}
\begin{gathered}
R_{2}(p) < R_{\rm crit}(p), \quad 0 < p < 1/2; \\
R_{0} < R_{2}(p) < R_{1}(p), \quad p < p_{0}; \qquad
R_{2}(p) = R_{1}(p) < R_{0}, \quad p > p_{0}; \\
R_{1}(p) < R_{\rm crit}(p), \quad p > p_{1}; \qquad
R_{1}(p) > R_{\rm crit}(p), \quad p < p_{1}; \\
R_{\rm crit}(p) \leq R_{0}, \quad p \geq 0,05014.
\end{gathered}
\end{equation}

In Fig. 1 plots of functions $R_{1}(p)$, $R_{2}(p)$,
$R_{\rm crit}(p)$ and $C(p)$ are shown.

{\it Remark} 1. Although notations $\omega_{R}(\alpha)$  and
$\omega_{1}(p)$ (also $\tau_{R}(\alpha)$ and $\tau_{1}(p)$) are not
well consistent, it should not imply any problems
(for example, we always have $R < 1$).

In the region $R_{\rm crit}(p) \leq R \leq C(p) = 1 -h_{2}(p)$
the function $E(R,p)$ is known since a long time ago
\cite{E} and it coincides with the sphere-packing bound
\begin{equation}\label{main21}
E(R,p) = E_{\rm sp}(R,p), \qquad  R_{\rm crit}(p) \leq R \leq C(p),
\end{equation}
where
\begin{equation}\label{spherepack}
\begin{gathered}
E_{\rm sp}(R,p) = D\left(\delta_{GV}(R)\| p\right), \qquad
D(x \| y) = x\log \frac{x}{y} + (1-x)\log \frac{1-x}{1-y}, \\
E_{\rm sp}(0,p) = \frac{1}{2}\log \frac{1}{4pq} = 2E(0,p).
\end{gathered}
\end{equation}

The main result of the paper is

{T h e o r e m \ 1}. 1) {\it For any $0 < p < 1/2$ the inequality
holds
\begin{equation}\label{main2}
E(R,p) = 1 - \log_{2}\left(1+2\sqrt{pq}\right) - R, \qquad
R_{2}(p) \leq R \leq R_{\rm crit}(p),
\end{equation}
where $R_{2}(p), R_{\rm crit}(p)$ are defined in \eqref{defR2} and
\eqref{defRcrit}, respectively}.

2) {\it For any $0 < p < 1/2$  and
$0 \leq R \leq \min\{R_{0},R_{2}(p)\}$ the bound is valid
\begin{equation}\label{defup1}
\begin{gathered}
E(R,p) \leq \frac{\omega_{R}}{2}
\log\frac{1}{4pq} -\mu(R,1/2,\omega_{R}) =
\frac{\omega_{R}}{2} \log \frac{1}{4pq} -
h_{2}(\tau) - h_{2}({\omega_{R}}) +1, \\
\omega_{R} = \frac{1}{2} - \sqrt{\tau(1-\tau)},
\qquad R = h_{2}(\tau),
\end{gathered}
\end{equation}
where $\omega_{R}$ и $\mu(R,1/2,\omega)$ are defined in \eqref{defGR}
and \eqref{mucor1}, respectively}.

3) {\it If $p < p_{0} \approx 0,036587$, then $R_{0} < R_{2}(p)$ and
for $R_{0} \leq R \leq R_{2}(p)$ the bound holds}
\begin{equation}\label{defup2}
\begin{gathered}
E(R,p) \leq 1 + \min_{0 \leq \alpha \leq 1/2}\left\{
\frac{G(\alpha,\tau)}{2} \log\frac{1}{4pq} -
L(G(\alpha,\tau))\right\} - R \leq \\
\leq 1 - R + \frac{\omega_{R}}{2} \log\frac{1}{4pq} -L(\omega_{R}),
\end{gathered}
\end{equation}
{\it where} ({\it $t_{1}(\omega)$ is  defined in \eqref{S.3}})
\begin{equation}\label{defL}
\begin{gathered}
L(\omega) = 2h_{2}\left[t_{1}(\omega)\right]-\omega -
(1-\omega)h_{2}\left[\frac{2t_{1}(\omega) - \omega}{2(1-\omega)}
\right].
\end{gathered}
\end{equation}

In other words, for any $0 < p < 1/2$ the function $E(R,p)$ is linear
for $R_{2}(p) \leq R \leq R_{\rm crit}(p)$. Earlier it was known only
for $R_{1}(p) \leq R \leq R_{\rm crit}(p)$, if $p$ is not too small
\cite{BM2,Bur6}. Recall that $R_{2}(p) < R_{1}(p)$, $p < p_{0}$, and
$R_{2}(p) = R_{1}(p)$, $p \geq p_{0}$. Also
$R_{2}(p) \leq \min\{R_{1}(p),R_{\rm crit}(p)\}$, $0 < p < 1/2$
(see also \eqref{F011}).

Notice also that improvement in the formula \eqref{main2} with respect
to \cite{Bur6} is attained due to using the values $\alpha < 1/2$
(in \cite{Bur6} only $\alpha = 1/2$ was used).

Inequalities \eqref{defup1}--\eqref{defup2} strengthen similar
estimate form \cite[теорема 1]{Bur6}.

For comparison purpose, the best know lowerbound for $E(R,p)$
has the form \cite{G1}
\begin{equation}\label{Elow1}
E(R,p) \geq -\delta_{GV}(R)\log(2\sqrt{pq}), \qquad
0 \leq R \leq R_{\rm min}(p),
\end{equation}
and
\begin{equation}\label{Elow2}
E(R,p) \geq 1 - \log_{2}\left(1+2\sqrt{pq}\right) - R, \qquad
 R_{\rm min}(p) \leq R \leq R_{\rm crit}(p),
\end{equation}
where $\delta_{GV}(R)$ is defined in \eqref{defR},
\begin{equation}\label{defRmin}
R_{\rm min}(p) = 1 - h_{2}\left(\frac{2\sqrt{pq}}
{1+2\sqrt{pq}}\right),
\end{equation}
and $R_{\rm min}(p) < R_{2}(p) < R_{\rm crit}(p)$, $0 < p < 1/2$.

Denote by $E_{\rm up}(R,p)$ the right-hand sides of formulas
\eqref{main2}--\eqref{defup2}, and by $E_{\rm low}(R,p)$
the right-hand sides of formulas
\eqref{Elow1}--\eqref{Elow2}. Then for all $R,p$ we have
$$
E_{\rm low}(R,p) \leq E(R,p) \leq E_{\rm up}(R,p).
$$

In Fig. 2 plots of functions $E_{\rm low}(R,p)$ and
$E_{\rm up}(R,p)$ for $p = 0,01$ are shown. In that case
$R_{\rm crit} \approx 0,5591$, $R_{1} \approx 0,5518$,
$R_{2} \approx 0,5370$, $R_{\rm min} \approx 0,3516$,
$C \approx 0,9192$.

Notice that $E_{\rm up}(R,p) - E_{\rm low}(R,p) > 0$ for
$R < R_{2}(p)$ and all $p$.

Upper bounds \eqref{defup1} and \eqref{defup2} have simple meaning:
we should apply the union bound using the best known upperbound
\eqref{MRRW} for the value $\delta(R)$ and the best known lowerbound
\eqref{spectr1} for the number of code neighbors.

When proving Theorem 1 we will need the function \cite{L1}:
\begin{equation}\label{defmu1}
\begin{gathered}
\mu(R,\alpha,\omega) = h_{2}(\alpha)- 2\int\limits_{0}^{\omega/2}
\log \frac{P + \sqrt{P^{2}-4Qy^{2}}}{Q}\,dy - (1-\omega)
h_{2}\left(\frac{\alpha - \omega/2}{1-\omega}\right), \\
P = P(y) = \alpha(1-\alpha) - \tau(1-\tau)-y(1-2y), \\
Q = Q(y) = (\alpha - y)(1-\alpha-y) =
\frac{(1-2y)^{2} - (1-2\alpha)^{2}}{4},
\end{gathered}
\end{equation}
where $\tau \leq 1/2$ such that $h_{2}(\tau) = h_{2}(\alpha)-1+R$.

Importance of the function $\mu (R,\alpha,\omega)$ and its relation
to the code spectrum $\{B_{i}\}$ (see \eqref{defBi}) is described
by the following variant \cite[Theorem 5]{L1} (see also proof in
\cite{Bur6}).

T h e o r e m \ 2. {\it For any $(R,n)$--code and any
$\alpha \in [\delta_{GV}(R),1/2]$ there exists
$\omega,\,0 \leq \omega \leq G(\alpha,\tau)$, where
$h_{2}(\tau) = h_{2}(\alpha)-1+R$ and
$G(\alpha,\tau)$ is defined in \eqref{defG}, such that
\begin{equation}\label{spectr1}
\frac{1}{n}\log B_{\omega n} \geq \mu(R,\alpha,\omega) + o(1),
\qquad n \to \infty,
\end{equation}
where $\mu(R,\alpha,\omega) > 0$ is defined in \eqref{defmu1} and
for $\mu(R,\alpha,\omega)$ the nonintegral representation
\eqref{reprmu1} holds}.

It should be noted that the parameter $\alpha$ determines a constant
weight $\alpha n$ code, which replaces the original code
(using Elias-Bassalygo lemma) \cite{M4,L1,Bur6}.

For $0 \leq R \leq R_{0}$ the best in Theorem 2 is $\alpha = 1/2$
\cite[Remark 4]{Bur5}, since such $\alpha$ simultaneously minimizes
$G(\alpha,\tau)$ and maximizes $\mu(R,\alpha,\omega)$ for all
 $\omega$. For $R > R_{0}$, probably, the optimal is
$\alpha = \alpha_{R}$ (see \eqref{defalphaR}), i.e. minimization of
$G(\alpha,\tau)$ over $\alpha < 1/2$ (at least, the value
$G(\alpha,\tau)$ is in the estimate \eqref{defup2}).
With $\alpha = \alpha_{R}$ we get from Theorem 2

C o r o l l a r y \ 1. {\it For any $(R,n)$--code there exists
$\omega$, $0 \leq \omega \leq \omega_{R}$ such that
\begin{equation}\label{spectr1a}
\frac{1}{n}\log B_{\omega n} \geq \mu(R,\alpha_{R},\omega) + o(1),
\qquad n \to \infty,
\end{equation}
where $\omega_{R}$ and $\alpha_{R}$ are defined in \eqref{defGR} and
\eqref{defalphaR}, respectively}.

{\it Remark} 2. Theorem 1 is based on the important feature of
inequalities \eqref{spectr1} and \eqref{spectr1a}. Till 1999 the best
upperbound for $E(R,p)$ followed from the best upperbound \eqref{MRRW}
for the maximal relative code distance
$\delta(R) \leq \omega_{R}$ \cite{M4}. From that point of view
inequalities \eqref{spectr1}--\eqref{spectr1a} do not improve
the estimate \eqref{MRRW}, but it follows from them that
$\mu(R,\alpha_{R},\omega_{R}) > 0$. In other words, there are an
exponential number of codewords on the minimal (or smaller) code
distance $\omega_{R}n$. That ``correction'' on
$\mu(R,\alpha_{R},\omega_{R})$ in inequalities
\eqref{defup1}--\eqref{defup2}, essentially, constitutes Theorem 1.
Notice also that the randomly chosen (``typical'') code has the
spectrum $n^{-1}\log B_{\omega n} \approx
h_{2}(\omega) - h_{2}(\delta_{GV}(R))$, $\omega \geq \delta_{GV}(R)$.
It is possible to check that for such code the maximal contribution
(additive) to the decoding error probability $P_{e}$ is given by
``neighbors'' on the distance $\omega_{1}(p)n$
(see \eqref{defomega1}), from which the inequality
\eqref{defRmin} follows for all $R,p$.

Introduce the value $R_{0}$ by the formula \cite{Lev1, Bur5}
\begin{equation}\label{defR0}
R_{0} = h_{2}(\tau_{0}) \approx 0,30524,
\end{equation}
where $\tau_{0} \approx 0,054507$ -- the unique root of the equation
\begin{equation}\label{eqR0}
(1-2\tau)\left[1 + \frac{1}{2\sqrt{\tau(1-\tau)}}\right] -
\ln \frac{1-\tau}{\tau} = 0.
\end{equation}

For fixed $R$ we have
\begin{equation}\label{dGda}
\frac{dG}{d\alpha} = \frac{2(1-2\alpha)}{1+2\sqrt{\tau(1-\tau)}} -
\frac{(1-2\tau)}{2\sqrt{\tau(1-\tau)}}\left[1 -
\frac{(1-2\alpha)^{2}}{\left(1+2\sqrt{\tau(1-\tau)}\right)^{2}}\right]
\frac{\ln[(1-\alpha)/\alpha]}{\ln[(1-\tau)/\tau]}.
\end{equation}

For any $R$ and $0 < \omega < G(\alpha,\tau)$ we also have
\cite[Proposition 1]{Bur5}
$$
\mu'_{\alpha}(R,\alpha,\omega) > 0, \qquad
\delta_{GV}(R) \leq \alpha < 1/2.
$$

For $R \leq R_{0}$ the function  $G(\alpha,\tau_{R}(\alpha))$
monotonically decreases in $\alpha \in [\delta_{GV}(R),1/2]$ and
$\alpha_{R} = 1/2$. If $R \in (R_{0},1)$ then values
$0 < \tau_{R} < \alpha_{R} < 1/2$ are uniquely defined by the system
of equations
\begin{equation}\label{syst1}
\begin{gathered}
\frac{dG}{d\alpha} = 0,
\end{gathered}
\end{equation}
\begin{equation}\label{syst1a}
h_{2}(\alpha)- h_{2}(\tau) = 1-R.
\end{equation}
Also
$$
\begin{gathered}
\tau_{R} = \frac{1}{2}\left\{1 - \sqrt{1-\left[
\sqrt{(1-\omega_{R})^{2}-(1-2\alpha)^{2}}- \omega_{R}\right]^{2}}
\right\}.
\end{gathered}
$$

For $\alpha \in (0,1/2)$ denote by $\tau(\alpha)$ the unique root of
the equation \eqref{syst1}, such that $\tau(\alpha) < \alpha$. Denote
also $R(\alpha) = 1 - h_{2}(\alpha) +
h_{2}(\tau(\alpha)),\; \alpha \in (0,1/2)$.

The function $\tau(\alpha)$ monotonically increases, and $R(\alpha)$
monotonically decreases on $\alpha$. Each $\alpha \in (0,1/2)$
defines $R(\alpha) > R_{0}$ and $\tau(\alpha) < \alpha$.
Similarly, each $R > R_{0}$ uniquely defines
$\alpha < 1/2$ and $\tau(\alpha) < \alpha$. Note that
$$
\lim_{\alpha \uparrow 1/2}\tau(\alpha) = \tau_{0} \approx 0,0545,
\quad \lim_{\alpha \uparrow 1/2}R(\alpha) = R_{0} \approx 0,305,
\quad \lim_{\alpha \uparrow 1/2}p(\alpha) = p_{0} \approx
0,036587.
$$

For calculation purpose it is convenient first to set the parameter
$\alpha \in (0,1/2)$, and then find sequentially corresponding
values $\tau(\alpha), R(\alpha)$ from \eqref{syst1}, \eqref{syst1a}
and $\omega(\alpha) = G(\alpha,\tau(\alpha))$. For $p = p(\alpha)$
(from \eqref{defR2}) we have $R_{2}(p) = R(\alpha)$ and
$$
p(\alpha) = \frac{1-G-\sqrt{1-2G}}{2(1-G)}, \qquad
G = G(\alpha,\tau(\alpha)).
$$

In particular, for $\alpha \to 0$ we have
$$
\begin{gathered}
\tau \approx \alpha^{2}, \quad G(\alpha,\tau(\alpha)) \approx \alpha,
\quad p(\alpha) \approx \frac{\alpha^{2}}{4}, \quad
C(p(\alpha)) \approx 1 - \frac{\alpha^{2}}{2}\log\frac{1}{\alpha}, \\
R_{2}(p(\alpha)) \approx 1 - \alpha\log\frac{1}{\alpha}, \quad
R_{\rm crit}(p(\alpha)) \approx 1 -
\frac{\alpha}{2}\log\frac{1}{\alpha}.
\end{gathered}
$$

Notice that due to Theorem 2 for a chosen $\alpha$ there exists
$\omega$ such that
$n^{-1}\log B_{\omega n} \geq \mu (R,\alpha,\omega) + o(1)$.
In other words, the number of neighbors on the distance $\omega n$
for each codeword $\boldsymbol{x}_{i}$ satisfies
{\it in average} that lowerbound. In fact, that property holds not
only {\it in average}, but also for {\it every} codeword
$\boldsymbol{x}_{i}$ from an ``essential'' part of all $M$
codewords (i.e. for $Me^{o(n)}$, $n \to \infty$ codewords). That
fact, established by the ``cleaning procedure'', regularly was used
in the papers \cite{Bur4, Bur5, Bur6} (and earlier) and will be also
used in the proof of Theorem 1.

{\it Remark} 3. In the author's paper \cite{Bur6} there are the
following inaccuracies:

3a) There is a miscalculation in the formula (15) \cite{Bur6} for
the function $\mu(R,1/2,\omega)$, coming from the earlier paper
\cite[formula (23)]{Bur5}. The correct version of that formula is
\cite[формула (23)]{Bur5}. Правильный вид этой формулы
\begin{equation}\label{mucor1}
\begin{gathered}
\mu(R,1/2,\omega) = - 2(1 - \omega)\log(1 - \omega)
- \log \tau - 2(1-\tau)\log(1-\tau) + \\
+ (1-2\tau)\log(\tau-\omega + g) + \log[1-\omega-(1-2\tau)g] -
2\omega\log g - 2,
\end{gathered}
\end{equation}
where
$$
\tau = \tau(R) = h_{2}^{-1}(R), \qquad g = g(\tau,\omega) =
\frac{1-2\tau+ \sqrt{(1-2\tau)^{2} - 4\omega(1-\omega)}}{2}.
$$

From \eqref{mucor1} the useful formula, which has already appeared
in \cite[formula (16)]{Bur6}, follows
\begin{equation}\label{mucor1/2}
\begin{gathered}
\mu(h_{2}(\tau),1/2,G(1/2,\tau)) = h_{2}(\tau) +
h_{2}(G(1/2,\tau)) - 1, \qquad \tau \geq 0.
\end{gathered}
\end{equation}

Due to importance of the formula \eqref{mucor1/2}, in Appendix
its derivation is presented. In \S 4 the explicit (non-integral)
representation for the function $\mu(R,\alpha,\omega)$ is obtained,
from which the formula \eqref{mucor1} can be received as well. There
also the generalization of the formula \eqref{mucor1/2} for
arbitrary $\alpha$ is obtained (see \eqref{ident1}).

2b) There is the following inaccuracy in the formulation of Theorem 1
\cite{Bur6}: the function $W(\omega,\alpha,R,p)$ was defined, using
the value $t_{2}(\omega,p)$ (for all $\omega$). In fact, the proof of
Theorem 1 in \cite{Bur6} was performed using the right function
$t(\omega,p)$ (and for that purpose the function $t(\omega,p)$ was
introduced in \cite[formula (29)]{Bur6}). Due to the author's fault,
the definition \cite[formula (9)]{Bur6}) (coming from the earlier
paper \cite{Bur5}) remained in \cite[\S 1]{Bur6}. Those changes do
not influence validity of the corollary 1 \cite{Bur6}.

2c) There is an inaccuracy in the proof of the Theorem 1 (noted by
Litsyn S., for which the author is grateful to him): the ``cleaning''
procedure in \cite[\S 4]{Bur6} was performed in such a way that
formally \cite[формула (44)]{Bur6} does not yet follow from
\cite[формула (41)]{Bur6}. The same drawback remained in
\cite{Bur6a} as well. In \S 2 below we fix that inaccuracy, and,
moreover, simplify the proof.

2d) The main difference of the paper with respect to \cite{Bur5}
is that here it turned out possible to investigate the case
$\alpha < 1/2$. An important role was played by the relation
\eqref{ident1}. Also some proof details were simplified.

In \S 2 connection between $P_{\rm e}$ and a code spectrum is
investigated. In \S 3 the proof of Theorem 1 is given. In \S 4 the
non-integral representation for $\mu(R,\alpha,\omega)$ is derived.
That representation is used in the proof of Theorem 1 in \S 3.
Some calculations and proofs are presented in Appendix.

\begin{center}
{\large\bf \S\;2. Lower bound for $P_{\rm e}$ and code spectrum}
\end{center}

For $\boldsymbol{x}_{i} \in {\cal C}$ denote
$d_{i}(\boldsymbol{y}) = d(\boldsymbol{x}_{i},\boldsymbol{y})$ and
for any integer $tn$, $t \in (0,1)$, introduce the set
$$
\begin{gathered}
{\mathbf X}_{t}(\boldsymbol{y}) = \left\{
\boldsymbol{x}_{i} \in {\cal C}: d_{i}(\boldsymbol{y}) = tn\right\} =
({\cal C} + \boldsymbol{y})^{(tn)}, \qquad \boldsymbol{y} \in F^{n}.
\end{gathered}
$$
Also for any integer $tn$, $t \in (0,1)$ and each pair of codewords
$\boldsymbol{x}_{i} \neq \boldsymbol{x}_{j}$ in the output space
$F^{n}$ of the channel introduce the ``ambiguity'' set:
$$
\begin{gathered}
{\mathbf Y}_{ij}(t)  = {\mathbf Y}_{ji}(t) =
\left\{\boldsymbol{y}:
d_{i}(\boldsymbol{y}) =d_{j}(\boldsymbol{y})= tn\right\}
\end{gathered}
$$
and for $i=1,\ldots,M$ the set
$$
\begin{gathered}
{\mathbf Y}_{i}(t) = \bigcup_{j \neq i} {\mathbf Y}_{ij}(t) =
\left\{\boldsymbol{y}: \mbox{there exists }
\boldsymbol{x}_{j} \neq \boldsymbol{x}_{i}, \
\mbox{such that } d_{i}(\boldsymbol{y}) =
d_{j}(\boldsymbol{y}) = tn\right\}.
\end{gathered}
$$

Next result is a variant of
\cite[Proposition 2, formulas (20), (21)]{Bur6} (see also
\cite[лемма 2]{Bur5}).

L e m m a \ 1. {\it For error probability $P_{e}$ the lower bound
holds} ($t = m/n$)
\begin{equation}  \label{nlow1aa}
\begin{gathered}
P_{e} \geq \frac{q^{n}}{2M}\sum_{m=1}^{n}
\left(\frac{p}{q}\right)^{m}\sum_{\mbox{\small\boldmath $y$}:
\left|{\mathbf X}_{t}(\mbox{\small\boldmath $y$})\right| \geq 2}
\left|{\mathbf X}_{t}(\boldsymbol{y})\right| =
\frac{q^{n}}{2M}\sum_{m=1}^{n}\left(\frac{p}{q}\right)^{m}\,
\sum_{i=1}^M \left|{\mathbf Y}_{i}(t)\right|.
\end{gathered}
\end{equation}

P r o o f. We explain only the equality in the formula
\eqref{nlow1aa} (it was not done in \cite{Bur6}). It is sufficient
to check the relation
$$
\sum_{\mbox{\small\boldmath $y$}:
|{\mathbf X}_{t}(\mbox{\small\boldmath $y$})| \geq 2}
|{\mathbf X}_{t}(\boldsymbol{y})| =
\sum_{i=1}^M \left|{\mathbf Y}_{i}(t)\right|, \qquad t = m/n > 0.
$$
For any point $\boldsymbol{y}$ with
$\left|{\mathbf X}_{t}(\boldsymbol{y})\right| \geq 2$ those
$\left|{\mathbf X}_{t}(\boldsymbol{y})\right|$ codewords
$\{\boldsymbol{x}_{i}\}$ give the same \\
contribution
$\left|{\mathbf X}_{t}(\boldsymbol{y})\right|$ to the right-hand
side of the equality. $\qquad \Box$

Since
$$
\begin{gathered}
\sum_{\mbox{\small\boldmath $y$}:\left|
{\mathbf X}_{t}(\mbox{\small\boldmath $y$})\right| \geq 2}
\left|{\mathbf X}_{t}(\boldsymbol{y})\right| =
\sum_{\mbox{\small\boldmath $y$}}
\left|{\mathbf X}_{t}(\boldsymbol{y})\right| -
\sum_{\mbox{\small\boldmath $y$}:\left|
{\mathbf X}_{t}(\mbox{\small\boldmath $y$})\right| = 1}
\left|{\mathbf X}_{t}(\boldsymbol{y})\right| = \\
= M\binom{n}{tn} -\left|\{\boldsymbol{y}:\left|
{\mathbf X}_{t}(\boldsymbol{y})\right| = 1\}\right|,
\end{gathered}
$$
then, in particular, from \eqref{nlow1aa} for any $t \in (0,1)$
we have
\begin{equation}\label{nlow1ab}
\begin{gathered}
P_{\rm e} \geq \frac{q^{n}}{2M}\left(\frac{p}{q}\right)^{tn}
\left[M\binom{n}{tn} - \left|\{\boldsymbol{y}:\left|
{\mathbf X}_{t}(\boldsymbol{y})\right| \geq 1\}\right|
\right].
\end{gathered}
\end{equation}
Using the inequality
$\left|\{\boldsymbol{y}:\left|{\mathbf X}_{t}
(\boldsymbol{y})\right| \geq 1\}\right|\leq 2^{n}$ and
optimizing over $t$, we get from \eqref{nlow1ab} the sphere-packing
bound (see \cite[\S 3]{Bur6})
\begin{equation}\label{spherepack1}
\begin{gathered}
E(R,p) \leq E_{\rm sp}(R,p), \qquad 0 \leq R \leq C(p),
\end{gathered}
\end{equation}
where $E_{\rm sp}(R,p)$ is defined in \eqref{spherepack}. From
the inequality \eqref{spherepack1} and similar lowerbound \cite{E}
we get the formula \eqref{main21}. In other words, for
$R \geq R_{\rm crit}(p)$ the lowerbound \eqref{nlow1aa} for $P_{e}$
is logarithmically precise.

Important for us will be the following result, similar to Johnson's
bound \\
\cite[Theorems 17.2.2 and 17.2.4]{MS} (see proof in Appendix).

{L e m m a \ 2}. {\it Let ${\mathcal C} =
\{\boldsymbol{x}_{1},\ldots,\boldsymbol{x}_{M}\}$ -- code of length
$n$ and constant weight $tn$, $t \leq 1/2$. If for some $\omega < 1/2$
and some $\delta \geq 0$ the following condition is fulfilled
\begin{equation}\label{adav1b}
\sum_{0 < i < \omega n}B_{i} \leq \delta M,
\end{equation}
and for some $a> 0$ the value $t$ satisfies the inequality
\begin{equation}\label{dav1aac}
t \leq \frac{1-\sqrt{1-2(1-\delta)\omega + 2a}}{2},
\end{equation}
then}
\begin{equation}\label{dav1abc}
M \leq \frac{\omega}{a}.
\end{equation}

In the sequel Lemma 2 will be used for small $\delta,a$. Then
\eqref{dav1aac} takes the form (see the definition of the value
$t_{1}(\omega)$ in \eqref{S.3})
$$
t \leq (1-\sqrt{1-2\omega})/2 + o(1), \qquad n \to \infty.
$$

Consider some values related to sums in the right-hand side of
\eqref{nlow1aa}. For codewords
$\boldsymbol{x}_{i},\boldsymbol{x}_{j}$ with
$d_{ij} = d(\boldsymbol{x}_{i},\boldsymbol{x}_{j}) = \omega n$
introduce the set
\begin{equation}\label{defZ}
{\mathbf Z}_{ij}(t,\omega) = \left\{\boldsymbol{y}:
d_{i}(\boldsymbol{y}) = d_{j}(\boldsymbol{y}) =tn\right\}.
\end{equation}
Since the cardinality $\left|{\mathbf Z}_{ij}(t,\omega)\right|$
does not depend on indices $(i,j)$, denote it simply $Z(t,\omega)$.
For the value $Z(t,\omega)$, $\omega/2 \leq t \leq 1/2$ we have
\begin{equation}\label{valZ}
\begin{gathered}
\frac{1}{n}\log_{2} Z(t,\omega) =
\frac{1}{n}\log_{2}\left[\binom{(1-\omega)n}{(t-\omega/2)n}
\binom{\omega n}{\omega n/2}\right] = u(t,\omega) -
\delta (t,\omega), \\
u(t,\omega) = \omega + (1-\omega)
h_{2}\left(\frac{2t-\omega}{2(1-\omega)}\right), \\
0 \leq \delta (t,\omega) \leq \frac{2}{n}\log_{2}\frac{n+2}{2},
\end{gathered}
\end{equation}
since for any $0 \leq k \leq n$ inequalities hold
\cite[formula (12.40)]{CT}
$$
\frac{1}{n+1}2^{nh(k/n)} \leq \binom{n}{k} \leq 2^{nh(k/n)}.
$$
For the function $u(t,\omega)$ we have
\begin{equation}\label{derc11}
\begin{gathered}
u'_{\omega} = -\frac{1}{2}\log\frac{(1-\omega)^{2}}
{(2t-\omega)(2-2t-\omega)} \leq 0, \\
u''_{\omega\omega} = -\frac{(1-2t)^{2}}
{(1-\omega)(2t-\omega)(2-2t-\omega)\ln 2} \leq 0, \\
u'_{t} = \log\frac{2-2t-\omega}{2t-\omega} \geq 0, \qquad
t \leq \frac{1}{2}, \\
u''_{tt} = \frac{4(1-\omega)}{(2t-\omega)(2-2t-\omega)\ln 2} \leq 0.
\end{gathered}
\end{equation}

We call  $(\boldsymbol{x}_{i},\boldsymbol{x}_{j})$ $\omega$--pair,
if $d_{ij} = \omega n$. Then the total number of $\omega$--pairs in
a code equals $MB_{\omega n}$. We say that a point $\boldsymbol{y}$
is $(\omega,t)$--covered, if there exists $\omega$--pair
$(\boldsymbol{x}_{i},\boldsymbol{x}_{j})$ such that
$d_{i}(\boldsymbol{y}) = d_{j}(\boldsymbol{y}) = tn$. For the point
$\boldsymbol{y}$ denote by $K(\boldsymbol{y},\omega,t)$ the number
of her $(\omega,t)$--coverings
(taking into account multiplicity of coverings), i.e.
\begin{equation}\label{defK}
K(\boldsymbol{y},\omega,t) =|\left\{(\boldsymbol{x}_{i},
\boldsymbol{x}_{j}): d_{ij} = \omega n,\,
d_{i}(\boldsymbol{y}) = d_{j}(\boldsymbol{y}) = tn\right\}|,
\qquad \omega > 0.
\end{equation}
Then for any $t, \boldsymbol{y}$
$$
|{\mathbf X}_{t}(\boldsymbol{y})|
\left(|{\mathbf X}_{t}(\boldsymbol{y})| - 1\right) =
\sum_{\omega > 0}K(\boldsymbol{y},\omega,t),
$$
and for any $t,\omega$ we get
\begin{equation}\label{defK2}
|{\mathbf X}_{t}(\boldsymbol{y})| \geq
\sqrt{K(\boldsymbol{y},\omega,t)}.
\end{equation}
Therefore from \eqref{nlow1aa} and \eqref{defK2} for any $t,\omega$
we have
\begin{equation}  \label{nlow1a1}
\begin{gathered}
P_{e} \geq \frac{q^{n}}{2M}\left(\frac{p}{q}\right)^{tn}
\sum_{\boldsymbol{y} \in {\mathbf Y}}
\sqrt{K(\boldsymbol{y},\omega,t)}.
\end{gathered}
\end{equation}

We modify the right-hand side of \eqref{nlow1a1} as follows. For
any set ${\cal A}$ denote
\begin{equation}\label{defK1}
K({\cal A},\omega,t) = \sum_{\boldsymbol{y} \in {\cal A}}
K(\boldsymbol{y},\omega,t),
\end{equation}
where $K(\boldsymbol{y},\omega,t)$ is defined in \eqref{defK}.
In other words, $K({\cal A},\omega,t)$ is the total number of
$(\omega,t)$--coverings of the set ${\cal A}$.

Introduce the set ${\mathbf Y}(\omega,t)$ of all
$(\omega,t)$--covered points $\boldsymbol{y}$, i.e.
$$
{\mathbf Y}(\omega,t) = \left\{\boldsymbol{y}:
K(\boldsymbol{y},\omega,t) \geq 1\right\} =\left\{\boldsymbol{y}:
\begin{array}{c}
\mbox{there exist } \boldsymbol{x}_{i},\boldsymbol{x}_{j}, \
\mbox{such that } \\
d_{ij} = \omega n \ \mbox{and }
d_{i}(\boldsymbol{y}) = d_{j}(\boldsymbol{y}) = tn
\end{array}
\right\}.
$$

Since every $\omega$--pair
$(\boldsymbol{x}_{i},\boldsymbol{x}_{j})$ $t$-covers
$Z(t,\omega)$ points $\boldsymbol{y}$, then
\begin{equation}\label{sumK}
K({\mathbf Y},\omega,t) = K({\mathbf Y}(\omega,t),\omega,t) =
MB_{\omega n}Z(t,\omega).
\end{equation}
For any subset ${\mathbf Y}' \subseteq {\mathbf Y}(\omega,t)$
introduce the value
\begin{equation}\label{defKmax}
K_{\rm max}({\mathbf Y}',\omega,t) =
\max_{\boldsymbol{y} \in {\mathbf Y}'}K(\boldsymbol{y},\omega,t).
\end{equation}
Then for any $t,\omega$ and
${\mathbf Y}' \subseteq {\mathbf Y}(\omega,t)$ из \eqref{nlow1a1}
we have
\begin{equation}  \label{nlow1a2}
\begin{gathered}
P_{e} \geq \frac{q^{n}}{2M}\left(\frac{p}{q}\right)^{tn}
\frac{K({\mathbf Y}',\omega,t)}
{\sqrt{K_{\rm max}({\mathbf Y}',\omega,t)}}.
\end{gathered}
\end{equation}

Describe the scheme of proving Theorem 1 realized in the paper.
Suppose that for chosen $\omega,t$ it is possible to choose also
a set ${\mathbf Y}'(\omega,t) \subseteq {\mathbf Y}(\omega,t)$
such that the following two conditions are fulfilled:
\begin{equation}\label{step1a}
K({\mathbf Y}'(\omega,t),\omega,t) \geq 2^{o(n)}
K({\mathbf Y}(\omega,t),\omega,t), \qquad n \to \infty
\end{equation}
and
\begin{equation}\label{step1b}
K(\boldsymbol{y},\omega,t) \leq 2^{o(n)}, \qquad
\boldsymbol{y} \in {\mathbf Y}'(\omega,t).
\end{equation}

Then the inequality \eqref{nlow1a2} can be continued as follows
\begin{equation}\label{idea1}
P_{\rm e} \geq 2^{o(n)}q^{n}\left(\frac{p}{q}\right)^{tn}
K({\mathbf Y}(\omega,t),\omega,t) = 2^{o(n)}q^{n}
\left(\frac{p}{q}\right)^{tn}B_{\omega n}Z(t,\omega).
\end{equation}
Estimate \eqref{idea1} is the desired additive lowerbound for
$P_{\rm e}$ (for values $\omega,t$). After \\
optimization of the
right-hand side of \eqref{idea1} over $\omega,t$ Theorem 1 will
be proved.

{\it Remark} 4. For a good code the value
$K(\boldsymbol{y},\omega,t)$ in \eqref{step1b}, probably, can not be
\\
exponential in $n$ for an essential part of all points
$\boldsymbol{y}$. In other words, for a good code it is unlikely that
exponential number of codewords are more probable than the true
codeword (!?).

We set some $\omega$ and $t = t(\omega)$. Due to \eqref{sumK}
there exists a collection of $M_{\omega}$ points
$\{\boldsymbol{y}_{1},\ldots,\boldsymbol{y}_{M_{\omega}}\}$
such that
$$
K(\boldsymbol{y}_{i},\omega,t) \sim
\frac{MB_{\omega n}Z(t,\omega)}{M_{\omega}}, \qquad
i = 1,\ldots,M_{\omega}.
$$
For that purpose it is sufficient to ``quantize'' all values
$\ln K(\boldsymbol{y},\omega,t) \sim n$ with a step of order $o(n)$,
$n \to \infty$. At that $K(\boldsymbol{y}_{i},\omega,t)$ is the
number of $\omega$-pairs on the $t$-sphere around the point
$\boldsymbol{y}_{i}$. The total number of various $\omega$-pairs
on $t$-spheres around points $\{\boldsymbol{y}_{i}\}$ has the order
$MB_{\omega n}$, i.e. "all" (in exponential sense) $\omega$-pairs
are located on those spheres. Therefore each $\omega$-pair belongs to
$Z(t,\omega)$ $t$-spheres, i.e. it covers $Z(t,\omega)$ points
$\boldsymbol{y}_{i}$. Then, essentially, each $\omega$-pair covers
only points $\boldsymbol{y}_{i}$.

If there are several such $M_{\omega}$ then choose the maximal one.

As a result, there are $N_{\omega}$ points on every $t$-sphere and
each $t$-sphere there are $K(\boldsymbol{y}_{i},\omega,t)$
$\omega$-pairs. We investigate the right-hand side of the inequality
\eqref{idea1}. Denoting
\begin{equation}\label{defbij}
\begin{gathered}
b(\omega) = \frac{1}{n}\log B_{\omega n} + o(1), \qquad
n \to \infty,
\end{gathered}
\end{equation}
represent \eqref{idea1} in an equivalent form ($n \to \infty$)
\begin{equation}\label{defF1a}
\begin{gathered}
\frac{1}{n}\log\frac{1}{P_{e}} \leq c(\omega,t,p) - b(\omega)+ o(1),
\end{gathered}
\end{equation}
where
\begin{equation}\label{defc}
\begin{gathered}
c(\omega,t,p) = t\log \frac{q}{p} - \log q - u(t,\omega),
\end{gathered}
\end{equation}
and the function $u(t,\omega)$ is defined in \eqref{valZ}. For the
function $c(\omega,t,p)$ using \eqref{derc11} we have
\begin{equation}\label{derc1}
\begin{gathered}
c'_{\omega} = -u'_{\omega} = \frac{1}{2}\log\frac{(1-\omega)^{2}}
{(2t-\omega)(2-2t-\omega)} \geq 0, \\
c''_{\omega\omega} = \frac{(1-2t)^{2}}
{(1-\omega)(2t-\omega)(2-2t-\omega)\ln 2} \geq 0, \\
c'_{t} = \log\frac{q(2t-\omega)}{p(2-2t-\omega)} \leq 0, \qquad
t \leq t_{2}(\omega,p) = \frac{\omega}{2} + (1-\omega)p, \\
c''_{tt} = \frac{4(1-\omega)}{(2t-\omega)(2-2t-\omega)\ln 2} \geq 0.
\end{gathered}
\end{equation}

The function $c(\omega,t,p)$ from \eqref{defc} has a simple meaning.
Suppose that we distinguish two codewords
$\boldsymbol{x}_{i},\boldsymbol{x}_{j}$ with
$d(\boldsymbol{x}_{i},\boldsymbol{x}_{j}) = \omega n$.
Introduce the set of ``ambiguity''
${\mathbf Z}_{ij}(t,\omega)$ from \eqref{defZ}. If
$\boldsymbol{y} \in {\mathbf Z}_{ij}(t,\omega)$ then with probability
$1/2$ decoding error occurs. Moreover,
$$
{\mathbf P}\left\{\boldsymbol{y} \in
{\mathbf Z}_{ij}(t,\omega)|\boldsymbol{x}_{i}\right\} \sim
2^{-c(\omega,t,p)n}.
$$

In order to choose the radius $t$ introduce functions
(see. \cite[formula (29)]{Bur6})
\begin{equation}\label{S.3}
t_{1}(\omega) = \dfrac{1-\sqrt{1-2\omega}}{2}, \qquad
t_{2}(\omega,p) = \frac{\omega}{2}+(1-\omega)p.
\end{equation}
The function $t_{2}(\omega,p)$ sometimes is called ``Elias radius''.
We set
\begin{equation}\label{deftn}
\begin{gathered}
t(\omega,p) = \min\{t_{1}(\omega),t_{2}(\omega,p)\} =
\left\{
\begin{array}{cc}
\left(1-\sqrt{1-2\omega}\right)/2, & \omega \leq \omega_{1}(p), \\
\omega/2 +(1-\omega)p, & \omega \geq \omega_{1}(p),
\end{array}
\right.
\end{gathered}
\end{equation}
where $\omega_{1}(p)$ is defined in \eqref{defomega1}. The
threshold value $\omega_{1}(p)$ will play very important role
in the sequel.

It follows from \eqref{derc1} that the function $c(\omega,t,p)$
monotonically decreases in $t < t_{2}(\omega,p)$ and monotonically
increases in $t > t_{2}(\omega,p)$. In particular,
\begin{equation}\label{minc}
\begin{gathered}
\min_{t}c(\omega,t,p) = c(\omega,t_{2}(\omega,p),p) =
\frac{\omega}{2}\log\frac{1}{4pq}.
\end{gathered}
\end{equation}

For any $\omega$ we will alway choose $t$ such that the following
condition is satisfied
\begin{equation}\label{constront}
\begin{gathered}
t \leq t(\omega,p).
\end{gathered}
\end{equation}
There are two reasons for such choice:

1) we would like to minimize the function $c(\omega,t,p)$, which
monotonically decreases in $t < t_{2}(\omega,p)$;

2) the condition $t \leq t_{1}(\omega)$ is necessary in order Lemma 2
be valid (and related with it the condition \eqref{step1b}).


\begin{center}
{\large\bf \S\;3. Proof of Theorem 1}
\end{center}

For a given $R$ choose some $\alpha$ such that
$h_{2}^{-1}(1-R) \leq \alpha \leq 1/2$ (see Theorem 2) and set
$\tau = h_{2}^{-1}(h_{2}(\alpha)-1+R)$ (such $\tau$ minimizes
$G(\alpha,\tau)$). Due to Theorem 3 for some
$\delta \leq \delta_{0} = G(\alpha,\tau)$ we have
$$
\frac{1}{n}\log B_{\delta n} \geq \mu(R,\alpha,\delta) + o(1),
\qquad n \to \infty.
$$

Since $K({\mathbf Y},\delta,s(\delta)) =
MB_{\delta n}Z(s(\delta),\delta)$ for any radius $s(\delta)$,
there exists a collection of $N_{\delta}$ points
$\{\boldsymbol{y}_{1},\ldots,\boldsymbol{y}_{N_{\delta}}\}$
such that
$$
K(\boldsymbol{y}_{i},\delta,s(\delta)) =
\frac{2^{o(n)}MB_{\delta n}Z(s(\delta),\delta)}{N_{\delta}}, \qquad
i = 1,\ldots,N_{\delta}.
$$
Therefore from \eqref{nlow1a1} we get
\begin{equation}  \label{nlow1a4}
\begin{gathered}
P_{e} \geq 2^{o(n)}\frac{q^{n}}{\sqrt{M}}
\left(\frac{p}{q}\right)^{s(\delta)n}
\sqrt{N_{\delta}B_{\delta n}Z(s(\delta),\delta)}.
\end{gathered}
\end{equation}

Now two cases are possible:

1) $\ln K(\boldsymbol{y}_{i},\delta,s(\delta)) = o(n)$ for an
essential part $N_{\delta}$ of points $\{\boldsymbol{y}_{i}\}$
(i.t. the condition \eqref{step1b} is fulfilled);

2) $\ln K(\boldsymbol{y}_{i},\delta,s(\delta)) \sim n$ for
for an essential part $N_{\delta}$ of points
$\{\boldsymbol{y}_{i}\}$.

Consider sequentially those cases assuming
$s(\delta) \leq t_{1}(\delta)$.

1) If  $\ln K(\boldsymbol{y}_{i},\delta,s(\delta)) = o(n)$ for an
essential part $N_{\delta}$ of points $\{\boldsymbol{y}_{i}\}$,
then the bound \eqref{nlow1a4} takes additive form for
$\delta$ and $t = s(\delta)$
\begin{equation}  \label{nlow1a5}
P_{\rm e} \geq 2^{o(n)}q^{n}\left(\frac{p}{q}\right)^{s(\delta)n}
B_{\delta n}Z(s(\delta),\delta).
\end{equation}

2) If $\ln K(\boldsymbol{y}_{i},\delta,s(\delta)) \sim n$ for an
essential part $N_{\delta}$ of points $\{\boldsymbol{y}_{i}\}$,
then consider $s(\delta)$-spheres around each point
$\boldsymbol{y}_{i}$ from that essential part. We may assume that
on every such $s(\delta)$-sphere there is the same number $m_{1}$
points and there are
$K(\boldsymbol{y}_{i},\delta,s(\delta))$ $\delta$-pairs. Denote by
$B_{\omega n}'$ analogues of numbers $B_{\omega n}$ for
$s(\delta)$-spheres. Then
$$
m_{1}B_{\delta n}' = K(\boldsymbol{y}_{i},\delta,s(\delta)) =
\frac{2^{o(n)}MB_{\delta n}Z(s(\delta),\delta)}{N_{\delta}}.
$$
Let for that essential part $N_{\delta}$ точек
$\{\boldsymbol{y}_{1},\ldots,\boldsymbol{y}_{N_{\delta}}\}$
the condition is satisfied
\begin{equation}\label{defprev2ca}
\sum_{\omega < \delta}K(\boldsymbol{y}_{i},\omega,s(\delta)) \leq
K(\boldsymbol{y}_{i},\delta,s(\delta))/n.
\end{equation}
Since $s(\delta) \leq t_{1}(\delta)$, it follows from Lemma 2
that the number points $m_{1}$ on every such $s(\delta)$-sphere
satisfies the inequality \eqref{dav1abc}, i.e. it is
non-exponential. Therefore in that case the additive bound
\eqref{nlow1a5} holds.

It remains to consider the case when for some $\omega < \delta$
the condition \eqref{defprev2ca} is not satisfied, i.e. for an
essential part $N_{\delta}$ points $\{\boldsymbol{y}_{i}\}$
\begin{equation}\label{defprev2d}
K(\boldsymbol{y}_{i},\omega,s(\delta)) >
K(\boldsymbol{y}_{i},\delta,s(\delta))/n^{2}.
\end{equation}
If necessary, choose the minimal one among all possible
$\omega < \delta$. Then we have
(since $K({\mathbf Y},\omega,s) = MB_{\omega n}Z(s,\omega)$)
\begin{equation}\label{defprev2g}
\log\left[B_{\omega n}Z(s(\delta),\omega)\right] \geq
\log\left[B_{\delta n}Z(s(\delta),\delta)\right] + o(n).
\end{equation}
Therefore for an essential part $N_{\delta}$ of points
$\{\boldsymbol{y}_{i}\}$ we have
$\log B_{\omega n}' \geq \log B_{\delta n}' + o(n)$. We choose
$s(\omega) \leq t_{1}(\omega)$ such that the potential additive
bound $\omega$ will be not less than the right-hand side of
\eqref{nlow1a5}, i.e. the inequality holds
\begin{equation}  \label{nlow1a6}
\left(\frac{p}{q}\right)^{s(\delta)n}
B_{\delta n}Z(s(\delta),\delta) \leq
\left(\frac{p}{q}\right)^{s(\omega)n}B_{\omega n}Z(s(\omega),\omega).
\end{equation}
Due to \eqref{defprev2g} for that purpose it is sufficient to have
$$
\left(\frac{p}{q}\right)^{s(\delta)n}Z(s(\delta),\omega) \leq
\left(\frac{p}{q}\right)^{s(\omega)n}Z(s(\omega),\omega),
$$
or, equivalently,
\begin{equation}\label{defprev2f}
\begin{gathered}
f = \left[s(\delta) - s(\omega)\right]\log\frac{q}{p} +
u(s(\omega),\omega) - u(s(\delta),\omega) \geq 0,
\end{gathered}
\end{equation}
where $u(t,\omega)$ is defined in \eqref{valZ}. Using \eqref{derc11}
we have
$$
\begin{gathered}
\frac{\partial f}{\partial s(\omega)} =
\log\frac{p[2-2s(\omega)-\omega]}{q[2s(\omega)-\omega]} \geq 0,
\qquad s(\omega) \leq t_{2}(p,\omega), \\
\frac{\partial f}{\partial s(\delta)} =
\log\frac{q[2s(\delta)-\omega]}{p[2-2s(\delta)-\omega]} \geq 0,
\qquad s(\delta) \geq t_{2}(p,\omega).
\end{gathered}
$$
Therefore for all $\omega$ we set
$$
s(\omega) = t(p,\omega) = \min\{t_{1}(\omega),t_{2}(p,\omega)\}.
$$
Since $s(\omega) \leq s(\delta)$, the inequalities \eqref{defprev2f}
and \eqref{nlow1a6} will be fulfilled. It means that a descent from
$\delta$ to $\omega$ does not decreases the potential additive
bound. It should be noted that a further descent from $\omega$ on
a lower level $\omega_{1}$ is not possible, since the level $\omega$
was chosen as the minimal possible one, for which the inequality
\eqref{defprev2d} holds. It proves validity of the additive bound
\eqref{nlow1a5} for $s(\omega) = t(p,\omega)$, from which we get
($h_{2}(\tau) = h_{2}(\alpha)-1+R$)

{P r o p o s i t i o n \ 1}. {\it For any $0 \leq R < C(p)$ and
$0 < p < 1/2$ the inequality holds}
\begin{equation}  \label{nlow1a8}
\begin{gathered}
E(R,p) \leq \min_{0 \leq \alpha \leq 1/2}
\max_{\delta \leq G(\alpha,\tau)}\left\{t(p,\delta)\log\frac{q}{p} -
\log q-\mu(R,\alpha,\delta) - u(t(p,\delta),\delta)\right\}.
\end{gathered}
\end{equation}

That result coincides with \cite[Theorem 1]{Bur6} in the most
interesting region $0 \leq R \leq R_{2}(p)$ and improves that
Theorem in the less interesting region $R_{2}(p) \leq R \leq C(p)$.

{\it Remark} 5. For large $R$ maximum over
$\delta \leq G(\alpha,\tau)$ in \eqref{nlow1a8} is attained not in
the extreme point $\delta = G(\alpha,\tau)$.

We will get the paper main Theorem 1 as a corollary from the bound
\eqref{nlow1a8}. For that purpose introduce the function
\begin{equation}\label{defWa}
W(\omega,\alpha,R,p) = \frac{\omega}{2} \log \frac{1}{4pq} -
\mu (R,\alpha,\omega).
\end{equation}

As will be clear below, it is sufficient to consider the case
$0 \leq R \leq R_{2}(p)$. Then
$\min\limits_{0 \leq \alpha \leq 1/2}
G(\alpha,\tau) \geq \omega_{1}(p)$ (see \eqref{defGR} and
\eqref{defR2}) and then $t(p,\delta) = t_{2}(\delta,p)$. Since
$$
u(t_{2}(\delta,p),\delta) = \delta + (1-\delta)h_{2}(p),
$$
then for $0 \leq R \leq R_{2}(p)$ we get from \eqref{nlow1a8}
\begin{equation}  \label{nlow1a9}
\begin{gathered}
E(R,p) \leq \min_{0 \leq \alpha \leq 1/2}
\max_{\delta \leq G(\alpha,\tau)}W(\delta,\alpha,R,p),
\end{gathered}
\end{equation}
where the function $W(\delta,\alpha,R,p)$ is defined in \eqref{defWa}.

We show that maximum over $\delta$ in the right-hand side of
\eqref{nlow1a9} is attained for $\delta = G(\alpha,\tau)$.

{L e m m a \ 3}. {\it For any $0 \leq \tau \leq \alpha \leq 1/2$ such
that $R = 1 - h_{2}(\alpha) + h_{2}(\tau)$ and
$G(\alpha,\tau) \geq \omega_{1}(p)$ the formula holds}
\begin{equation}\label{cond19}
\begin{gathered}
\max_{\lambda \leq G(\alpha,\tau)}W(\lambda,\alpha,R,p) =
W(G(\alpha,\tau),\alpha,R,p) = \\
= \frac{G(\alpha,\tau)}{2} \log \frac{1}{4pq} -
\mu (R,\alpha,G(\alpha,\tau)).
\end{gathered}
\end{equation}

{P r o o f}.
From \eqref{defmu1} we have \cite[Proposition 1]{Bur5}
($P = P(\omega/2)$, $Q = Q(\omega/2)$,
$a_{1} = 2[\alpha(1-\alpha)-\tau(1-\tau)]$)
\begin{equation}\label{defmu12}
\begin{gathered}
\mu'_{\omega}(R,\alpha,\omega) = \frac{1}{2}\log\frac{(1-\omega)^{2}}
{(\alpha-\omega/2)(1-\alpha-\omega/2)} -
\log\frac{P + \sqrt{P^{2} - Q\omega^{2}}}{Q} =  \\
= \log\frac{(1-\omega)\sqrt{(2\alpha-\omega)(2-2\alpha-\omega)}}
{a_{1}-\omega(1-\omega) +
\sqrt{(1-2\tau)^{2}\omega^{2} -2a_{1}\omega + a_{1}^{2}}}, \\
\mu''_{\omega \omega}(R,\alpha,\omega) < 0.
\end{gathered}
\end{equation}
For $W(\omega,\alpha,R,p)$ we have from \eqref{defWa} and
\eqref{defmu12} \cite[Proposition 2]{Bur5}
$$
\begin{gathered}
W''_{\omega \omega}(\omega,\alpha,R,p) < 0, \qquad
W'_{\omega}(\omega,\alpha,R,p)\Big|_{\omega = G} =
\log\frac{G}{\sqrt{4pq}(1-G)},  \qquad G = G(\alpha,\tau).
\end{gathered}
$$
Therefore, if $G(\alpha,\tau) \geq \omega_{1}(p)$ then
$W'_{\omega}(\omega,\alpha,R,p)\Big|_{\omega = G} \geq 0$, and then
maximum over $\delta$ in the right-hand side of \eqref{cond19} is
attained for $\lambda = G(\alpha,\tau)$. $\qquad \Box$

As a result, from \eqref{nlow1a9} and \eqref{cond19} for
$0 \leq R \leq R_{2}(p)$ we get
\begin{equation}  \label{nlow1a10}
\begin{gathered}
E(R,p) \leq \min_{0 \leq \alpha \leq 1/2}\left\{
\frac{G(\alpha,\tau)}{2} \log \frac{1}{4pq} -
\mu (R,\alpha,G(\alpha,\tau))\right\}.
\end{gathered}
\end{equation}

It remained us to get for $\mu(R,\alpha,G(\alpha,\tau))$ from
\eqref{nlow1a10} an explicit expression. We use the following
analytical result (see proof in Appendix).

{L e m m a \ 4}. {\it For any $\alpha,\tau$ the formula holds}
\begin{equation}\label{ident1}
\begin{gathered}
\mu(R,\alpha,G(\alpha,\tau)) = L(G(\alpha,\tau)) + R -1,
\end{gathered}
\end{equation}
{\it where $R = 1 - h_{2}(\alpha) + h_{2}(\tau)$ and the function
$L(\omega)$ is defined in \eqref{defL}}.

Using \eqref{ident1} and \eqref{nlow1a10} we get

{П р е д л о ж е н и е \ 2}. {\it For any $0 < p < 1/2$ and
$0 \leq R \leq R_{2}(p)$ the bound holds
\begin{equation}  \label{nlow1a11}
\begin{gathered}
E(R,p) \leq 1 - R + \min_{0 \leq \alpha \leq 1/2}\left\{
\frac{G(\alpha,\tau)}{2} \log \frac{1}{4pq} -
L(G(\alpha,\tau))\right\} \leq \\
\leq 1 - R + \frac{\omega_{R}}{2} \log\frac{1}{4pq} -L(\omega_{R}),
\end{gathered}
\end{equation}
where $\omega_{R}$ is defined in \eqref{defGR}}.

In particular, from \eqref{nlow1a11} the formula \eqref{defup2}
follows. Concerning the formula \eqref{defup1} recall that if
$R \leq R_{0}$, then the best is $\alpha = 1/2$. The bound
\eqref{defup1} follows from \eqref{nlow1a10} with
$\alpha = 1/2$ and \eqref{mucor1/2}.

It remained us to prove the formula \eqref{main2}. Note that
$$
t_{1}(\omega_{1}(p)) = \frac{\sqrt{p}}{\sqrt{q} + \sqrt{p}}.
$$
Therefore if $G(\alpha,\tau) = \omega_{1}(p)$ then using
\eqref{ident1} we get
\begin{equation}\label{cond1bc}
\begin{gathered}
\mu(R,\alpha,\omega_{1}(p)) =
\frac{\omega_{1}(p)}{2}\log\frac{1}{4pq} + R + \log(1+2\sqrt{pq}) -1.
\end{gathered}
\end{equation}
Then for any rate $R$, for which it is possible to have
$G(\alpha,\tau) = \omega_{1}(p)$, from \eqref{nlow1a10} and
\eqref{cond1bc} the inequality follows
\begin{equation}\label{part11b}
\begin{gathered}
E(R,p) \leq 1 - \log(1+2\sqrt{pq}) - R.
\end{gathered}
\end{equation}
The rate $R = R_{2}(p)$ is the minimal of such rates
(see \eqref{defR2}). For $R = R_{\rm crit}(p)$ the formula holds
\cite{E} (see \eqref{main21})
\begin{equation}\label{Esp2}
\begin{gathered}
E\left(R_{\rm crit},p\right) =
E_{\rm sp}\left(R_{\rm crit},p\right) =
1 - \log_{2}\left(1+2\sqrt{pq}\right) - R_{\rm crit}.
\end{gathered}
\end{equation}
Therefore due to to the ``straight-line upper bound'' \cite{SGB1}
the inequality \eqref{part11b} holds for all $R$ such that
$R_{2}(p) \leq R \leq R_{\rm crit}(p)$, i.e.
\begin{equation}\label{part11a}
\begin{gathered}
E(R,p) \leq 1 - \log(1+2\sqrt{pq}) - R, \qquad
R_{2}(p) \leq R \leq R_{\rm crit}(p).
\end{gathered}
\end{equation}

On the other hand, for the function $E(R,p)$ the random coding
lower bound is known \cite{E}
\begin{equation}\label{gen0a}
E(R,p) \geq 1 - \log_{2}\left(1+2\sqrt{pq}\right) - R, \qquad
0 \leq R \leq R_{\rm crit}(p).
\end{equation}
As result, from \eqref{part11a} and \eqref{gen0a} the formula
\eqref{main2} follows, that completes Theorem 1 proof.


\begin{center}
{\large\bf \S\;4. Non-integral representation for
$\mu(R,\alpha,\omega)$}
\end{center}

Proof of the next result represents a standard integration
using Euler's substitution. That representation is used in
deriving the formula \eqref{ident1} and in the proof of Theorem 1.

P r o p o s i t i o n \ 3. {\it For the function
$\mu(R,\alpha,\omega)$ the representation holds
\begin{equation}\label{reprmu1}
\begin{gathered}
\mu(R,\alpha,\omega) =(1-\omega)h_{2}\left(\frac{\alpha- \omega/2}
{1-\omega}\right) -h_{2}(\alpha) + 2h_{2}(\omega) +
\omega\log\frac{2\omega}{e} - T(A,B,\omega),
\end{gathered}
\end{equation}
where
\begin{equation}\label{auxdefmu0}
\begin{gathered}
\tau = h_{2}^{-1}(h_{2}(\alpha)-1+R) \leq 1/2, \\
A = 1-2\alpha, \qquad B = 1 -2\tau, \qquad B > A \geq 0, \\
T(A,B,\omega) = \omega\log(v-1)-
(1-\omega)\log\frac{v^{2} - A^{2}}{v^{2} - B^{2}} + \\
+ B\log\frac{v+B}{v-B}- A\log\frac{v+A}{v-A} -
\frac{(v-1)(B^{2} - A^{2})}{(v^{2} - B^{2})\ln 2},
\end{gathered}
\end{equation}
and }
\begin{equation}\label{auxdefmu01}
\begin{gathered}
v = \frac{\sqrt{B^{2}\omega^{2} - 2a_{1}\omega + a_{1}^{2}} +
a_{1}}{\omega}, \qquad
a_{1} = 2[\alpha(1-\alpha)-\tau(1-\tau)].
\end{gathered}
\end{equation}

P r o o f. Using notations \eqref{auxdefmu0} and the variable
$z = 2y$, we have from \eqref{defmu1}
$$
\begin{gathered}
\mu (R,\alpha,\omega) = h_{2}(\alpha) - \omega -
(1-\omega)h_{2}\left(\frac{\alpha - \omega/2}{1-\omega}\right) -
\int\limits_{0}^{\omega}\log\frac{f_{1}}{g_{1}}\,dz,\\
f_{1} = z^{2} - z+a_{1}+\sqrt{B^{2}z^{2}-2a_{1}z+a_{1}^{2}},
\qquad g_{1} = (1-z)^{2} - A^{2}.
\end{gathered}
$$
Then
$$
\begin{gathered}
\int\limits_{0}^{\omega}\log g_{1}\,dz = - 2h_{2}(\alpha) -
2(1-\omega)\log(1-\omega) + 2(1-\omega)h_{2}
\left(\frac{\alpha - \omega/2}{1-\omega}\right) -
2\omega\log\frac{e}{2}.
\end{gathered}
$$
We also have
$$
\begin{gathered}
\int\limits_{0}^{\omega}\log f_{1}\,dz  =
F(a_{1},B,\omega) - F(a_{1},B,0),
\end{gathered}
$$
where $F(a_{1},B,z)$ -- primitive function for $\log f_{1}$.
In order to find $F(a_{1},B,z)$, we use Euler's substitution
$$
\begin{gathered}
\sqrt{B^{2}z^{2} - 2a_{1}z + a_{1}^{2}} = zv - a_{1},
\end{gathered}
$$
and then
$$
\begin{gathered}
z = \frac{2a_{1}(v-1)}{v^{2}-B^{2}}, \qquad
z' = \frac{2a_{1}(2v-v^{2}-B^{2})}{(v^{2}-B^{2})^{2}}.
\end{gathered}
$$
Now
$$
\begin{gathered}
F(a_{1},B,z)\ln 2 = \int\ln f_{1}\,dz = z\ln z - z + \\
+ z(v)\ln[z(v) + v - 1] +
\frac{a_{1}}{A}\ln\frac{v+A}{v-A} - 4a_{1}^{2}I_{1},
\end{gathered}
$$
where
$$
\begin{gathered}
I_{1} = \int\frac{(2v-v^{2}-B^{2})}
{(v^{2} -A^{2})(v^{2}-B^{2})^{2}}\,dv.
\end{gathered}
$$
After standard integration we have
$$
\begin{gathered}
(B^{2} - A^{2})^{2}I_{1} = \\
= \ln\frac{v^{2} - A^{2}}{v^{2} - B^{2}} +
\frac{(v-1)(B^{2} - A^{2})}{(v^{2} - B^{2})} +\frac{(B^{2} + A^{2})}
{2A}\ln\frac{v+A}{v-A} + B\ln\frac{v-B}{v+B}.
\end{gathered}
$$
Since $2a_{1} = B^{2} - A^{2}$ and
$$
z(v) + v- 1 = \frac{(v-1)(v^{2} - A^{2})}{v^{2} - B^{2}},
$$
we get
$$
\begin{gathered}
F(a_{1},B,z) = z\log z - \frac{z}{\ln 2} + z\log(v-1)-
(1-z)\log\frac{v^{2} - A^{2}}{v^{2} - B^{2}} + \\
+ B\log\frac{v+B}{v-B} - A\log\frac{v+A}{v-A} -
\frac{(v-1)(B^{2} - A^{2})}{(v^{2} - B^{2})\ln 2},
\end{gathered}
$$
where $v$ is defined in \eqref{auxdefmu01}. Since
$v \to \infty$ as $u \to 0$ и $F(a,b,0) = 0$, then
$$
\begin{gathered}
\mu(R,\alpha,\omega) =
-h_{2}(\alpha) - 2(1-\omega)\log (1-\omega) - 2\omega\log e + \\
+ (1-\omega)h_{2}\left(\frac{\alpha - \omega/2}{1-\omega}\right)
- F(a_{1},B,\omega),
\end{gathered}
$$
from which Proposition 3 follows.  $\qquad \Box$

\bigskip

\hfill {\large\sl APPENDIX}

\medskip

P r o o f \ o f \ f o r m u l a \ \eqref{mucor1}.
For $\omega \leq G(1/2,\tau)$ we have \cite[formula (П. 2)]{Bur5}
$$
\begin{gathered}
\mu(R,1/2,\omega)\ln 2 =
 -2\omega (1-\ln 2) -2(1 - \omega)\ln(1 - \omega) - I, \\
I = \int\limits_{0}^{2\omega}\ln\left[b + \sqrt{(1-z)^{2}+
b^{2}- 1}\right]\,dz, \qquad b = 1 -2\tau > 0.
\end{gathered}
$$
In order to find the integral $I$, we use variable
$u = \sqrt{(1-z)^{2}+b^{2}- 1}$. Then denoting
$$
A = \sqrt{(1-2\omega)^{2}+b^{2}- 1}, \qquad
v_{1} = \frac{b}{\sqrt{1-b^{2}}}, \qquad v_{2} =
\frac{A}{\sqrt{1-b^{2}}},
$$
and using integration by parts we have
$$
\begin{gathered}
I = - \int\limits_{b}^{A}\ln(b+u)\,d\sqrt{u^{2} +1-b^{2}} = \\
= \ln(2b) - (1-2\omega)\ln(b+A) +
\sqrt{1-b^{2}}\int\limits_{v_{1}}^{v_{2}}
\frac{\sqrt{1+v^{2}}}{v+v_{1}}\,dv.
\end{gathered}
$$
Since
$$
\begin{gathered}
\int\frac{\sqrt{1+z^{2}}}{z+a}\,dz =  \\
= \sqrt{1+z^{2}} - a\ln\left[z+\sqrt{1+z^{2}}\right] -
\sqrt{1+a^{2}}\,\ln\frac{\sqrt{(1+a^{2})(1+z^{2})}-za+1}{z+a},
\end{gathered}
$$
then
$$
\begin{gathered}
I = \ln(2b) - (1-2\omega)\ln(b+A) - 2\omega -
b\ln\frac{1-2\omega + A}{1+b} - \ln\frac{b(2-2\omega-b^{2}-bA)}
{(b+A)(1-b^{2})}.
\end{gathered}
$$
After standard algebra with $g = (b+A)/2$ we get the formula
\eqref{mucor1}. $\qquad \Box$

P r o o f \ o f \ l e m m a \ 2. Consider a code ${\mathcal C}$
average distance
$$
d_{\rm av}\left({\mathcal C}\right) =
M^{-2}\sum_{i=1}^{M}\sum_{j=1}^{M}d_{ij}.
$$
Similarly to Plotkin's bound derivation \cite[теорема 2.2.1]{MS}
we have
\begin{equation}\label{Plot1}
d_{\rm av}\left({\mathcal C}\right) \leq 2t(1-t)n, \qquad
0 \leq t \leq 1.
\end{equation}
Using the assumption \eqref{adav1b} we can also lower bound
the value $d_{\rm av}\left({\mathcal C}\right)$
$$
\begin{gathered}
d_{\rm av}\left({\mathcal C}\right) \geq \frac{1}{M}
\sum_{i \geq \omega n}iB_{i} \geq \frac{\omega n}{M}
\sum_{i \geq \omega n}B_{i} =  \\
= \frac{\omega n}{M}\left\{\sum_{i > 0}B_{i} -
\sum_{0 < i < \omega n}B_{i}\right\} \geq
\frac{\omega(M-1-\delta M)n}{M}.
\end{gathered}
$$
Comparing that estimate with \eqref{Plot1} we get the inequality
\eqref{dav1abc}. $\qquad \Box$

P r o o f \ o f \ f o r m u l a \ \eqref{ident1}. We use the
representation \eqref{reprmu1} and notations from
\eqref{auxdefmu0} and \eqref{auxdefmu01}. Since
$$
G(\alpha,\tau) = \frac{a_{1}}{1+2\sqrt{\tau(1-\tau)}}\,, \qquad
a_{1} = 2[\alpha(1-\alpha)-\tau(1-\tau)],
$$
then
$$
B^{2}G^{2} - 2a_{1}G + a_{1}^{2} = 0 \qquad \mbox{и} \qquad
v = \frac{a_{1}}{G} = 1+2\sqrt{\tau(1-\tau)}.
$$
Also
$$
\begin{gathered}
v-B = 2\sqrt{\tau v}\,, \qquad v+B = 2\sqrt{(1-\tau) v}, \\
v^{2} - B^{2} = 4\sqrt{\tau(1-\tau)}
\left[1+2\sqrt{\tau(1-\tau)}\right], \qquad
\frac{(v-1)(B^{2}-A^{2})}{v^{2}-B^{2}} = G, \\
v - A = 2\left[\sqrt{\tau(1-\tau)} + \alpha\right], \qquad
v + A = 2\left[1+\sqrt{\tau(1-\tau)} - \alpha\right].
\end{gathered}
$$
Therefore ($\omega = G(\alpha,\tau)$)
$$
\begin{gathered}
T(A,B,G)= \frac{\omega}{2}\log\left[\tau(1-\tau)\right] -
(1-\omega)\log\frac{\left[1+\sqrt{\tau(1-\tau)} - \alpha\right]
\left[\sqrt{\tau(1-\tau)} + \alpha\right]}
{\sqrt{\tau(1-\tau)}\left[1+2\sqrt{\tau(1-\tau)}\right]} + \\
+ \frac{(1-2\tau)}{2}\log\frac{1-\tau}{\tau} -
(1-2\alpha)\log\frac{1+\sqrt{\tau(1-\tau)} - \alpha}
{\sqrt{\tau(1-\tau)}+ \alpha} - \omega\left(\frac{1}{\ln 2}-1\right)
\end{gathered}
$$
and
$$
\begin{gathered}
\mu(R,\alpha,G) =
(1-\omega)h_{2}\left(\frac{\alpha- \omega/2}{1-\omega}\right) +
R - 1 +2h_{2}(\omega) + \omega\log \omega + \\
+ (1-\omega)\log\frac{\left[1+\sqrt{\tau(1-\tau)} - \alpha\right]
\left[\sqrt{\tau(1-\tau)} + \alpha\right]}
{\left[1+2\sqrt{\tau(1-\tau)}\right]} + \\
+ (1-2\alpha)\log\frac{1+\sqrt{\tau(1-\tau)} - \alpha}
{\sqrt{\tau(1-\tau)} + \alpha}.
\end{gathered}
$$
Using in the last expression formulas
$$
\begin{gathered}
\frac{2\alpha - G}{2} =
\frac{\left[\sqrt{\tau(1-\tau)} + \alpha\right]^{2}}
{1+2\sqrt{\tau(1-\tau)}}, \\
 1-G- \frac{2\alpha - G}{2} =
\frac{\left[1+\sqrt{\tau(1-\tau)} - \alpha\right]^{2}}
{1+2\sqrt{\tau(1-\tau)}},
\end{gathered}
$$
and also formulas
$$
\begin{gathered}
ah\left(b/a\right) = a\ln a - b\ln b - (a-b)\ln (a-b), \\
2t_{1}(\omega) - \omega = 2t_{1}^{2}(\omega),
\end{gathered}
$$
we get the formula \eqref{ident1}. $\qquad \Box$

\bigskip


\vspace{5mm}

\begin{flushleft}
{\small {\it Burnashev Marat Valievich} \\
Kharkevich Institute for Information Transmission Problems, \\
Russian Academy of Sciences, Moscow\\
 {\tt burn@iitp.ru}}
\end{flushleft}%

\newpage

\includegraphics[width=0.9\hsize,height=0.8\hsize]{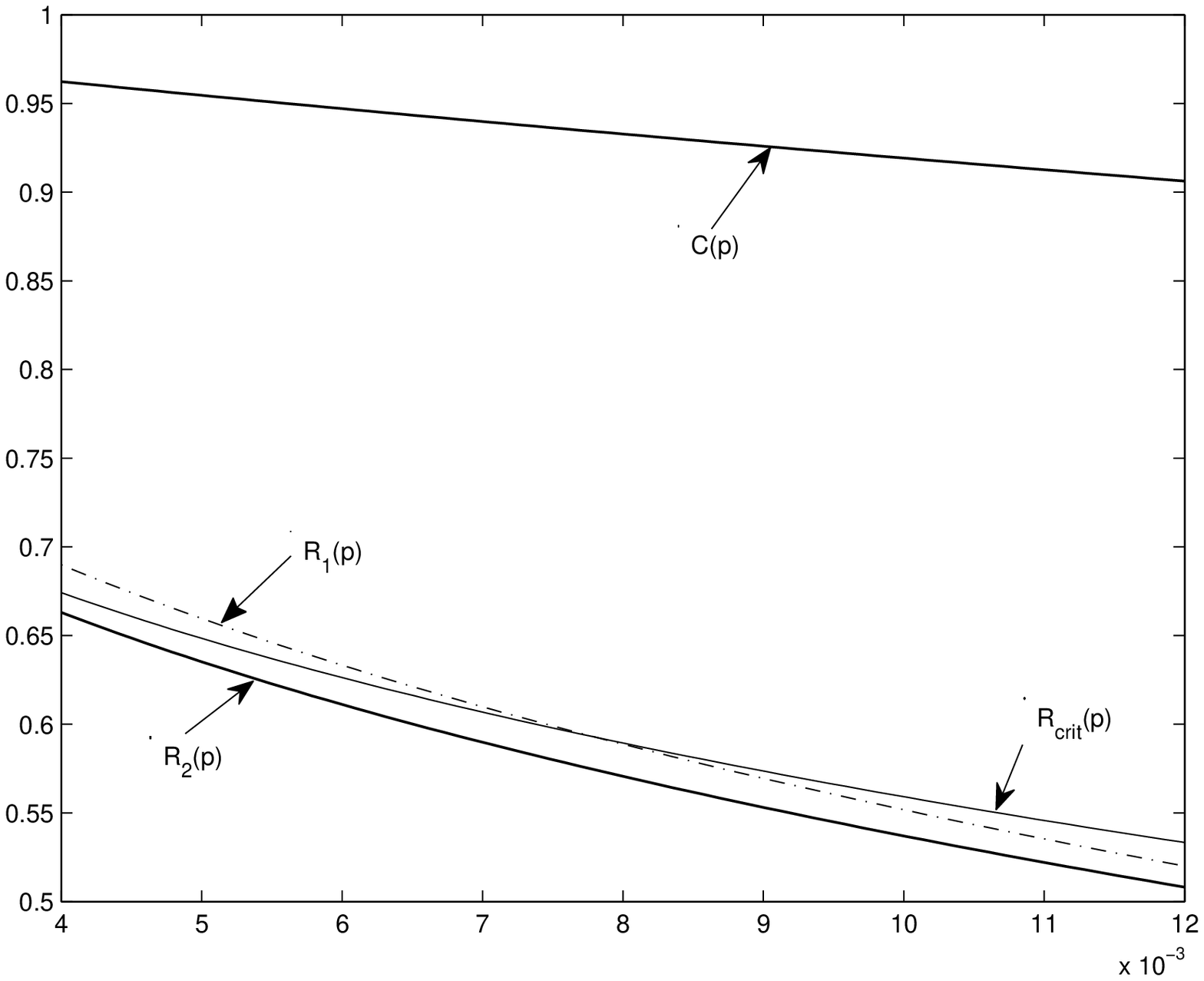}

\begin{center}
{ Fig. 1. Plots of functions $R_{1}(p),R_{2}(p),R_{\rm crit}(p)$
and $C(p)$}
\end{center}

\newpage

\includegraphics[width=0.9\hsize,height=0.8\hsize]{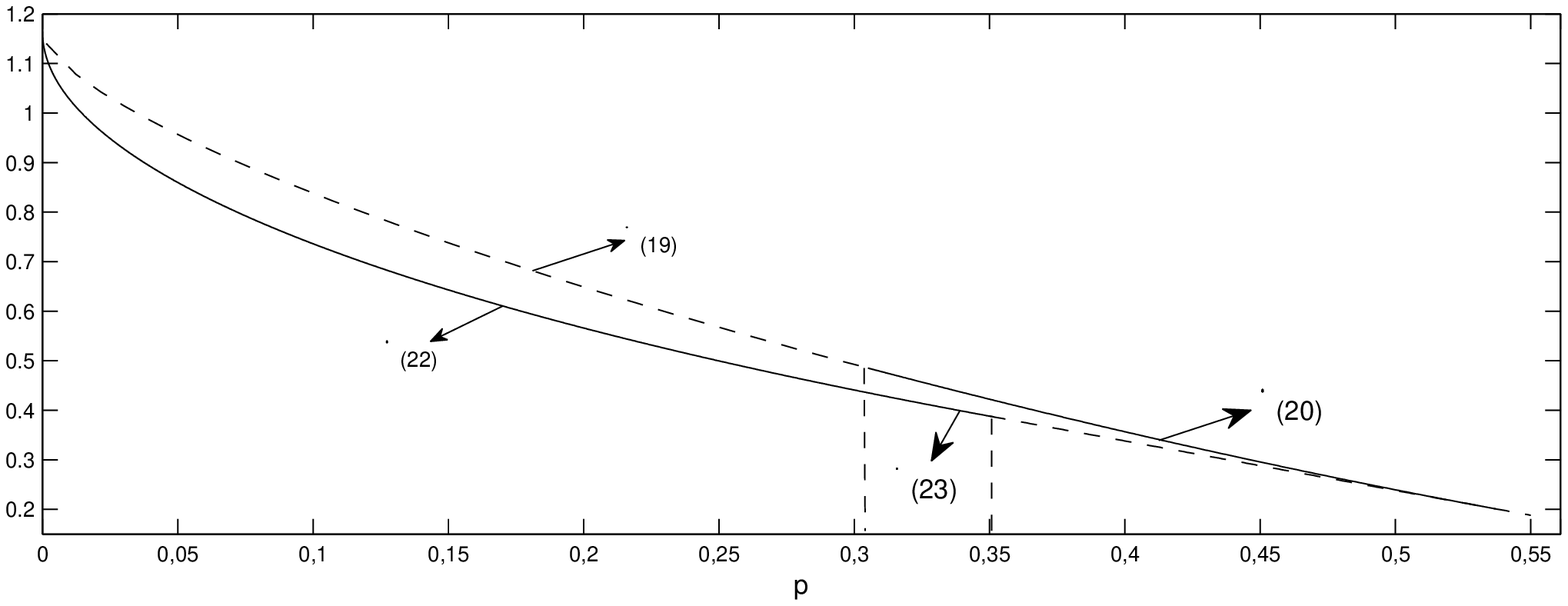}

\begin{center}
{ Fig. 2. Plots of functions $E_{\rm low}(R,p)$ and
$E_{\rm up}(R,p)$ for $p = 0,01$}
\end{center}

\end{document}